\documentclass[showkeys,showpacs,preprintnumbers,aps,twocolumn]{revtex4}

\usepackage{bm}
\usepackage[english]{babel}
\usepackage[latin1]{inputenc}
\usepackage[dvips]{graphicx}
\usepackage{amsmath}
\usepackage{amsfonts}

\usepackage{epsfig}

\begin{document}
\parindent = 0pt

\preprint{(2003)}

\date{\today}
\title{Shot Noise of coupled Semiconductor Quantum Dots}

\author{G.~Kie{\ss}lich}
\email{kieslich@physik.tu-berlin.de}
\thanks{Fax: +49-(0)30-314-21130}
\author{A.~Wacker}
\author{E.~Sch{\"o}ll}
\affiliation{Institut f{\"u}r Theoretische Physik, Technische Universit{\"a}t
  Berlin, D-10623 Berlin, Germany}

\begin{abstract}

The low-frequency shot noise properties of two electrostatically coupled semiconductor 
quantum dot states which are connected
to emitter/collector contacts are studied. A master equation approach is used to analyze 
the bias voltage 
dependence of the Fano factor as a measure of temporal correlations in tunneling current caused by 
Pauli's exclusion principle and the Coulomb interaction.
In particular, the influence of the Coulomb interaction on the shot noise behavior
is discussed in detail and predictions for future experiments will be given. 
Furthermore, we propose 
a mechanism for 
negative differential conductance and investigate the related super-Poissonian shot noise.

\end{abstract}

\maketitle


\section{Introduction}
Shot noise investigations in mesoscopic systems can reveal information of 
transport properties which are not accessible by conductance measurements alone \cite{BLA00}. 
In particular, the dynamic correlations in the tunneling current through double-barrier structures 
caused by Pauli's exclusion principle can provide information
regarding the barrier geometry \cite{CHE91}.
If the charging energy of bound states becomes larger than the thermal energy as in the case of small quantum dots (QDs), 
strong Coulomb correlations occur and have an additional influence on the shot noise. For metallic QDs the 
zero-frequency Fano factor which quantifies correlations with respect to the uncorrelated Poissonian noise \cite{SCH18}
was analyzed by a master equation approach including the Coulomb blockade effect in Ref.~\cite{HER93}. At the steps 
of the resulting Coulomb staircase the Fano factor shows dips caused by Coulomb correlations which is quantitatively  
confirmed in the experiment \cite{BIR95}. In Ref.~\cite{HAN93} a similar theoretical approach was applied 
to determine the finite-frequency shot noise of metallic QDs.

Despite that, for semiconductor QDs a comprehensive picture of the bias dependent shot noise behavior and a subsequent 
comparison with experimental data is not available at the moment. Some theoretical work has been 
done: the investigation of the Fano factor of an ensemble of states with statistically varying
positions in a barrier by a classical approach \cite{NAZ96}; analytical bias dependence of the Fano factor
for zero-temperature with non-equilibrium Green's functions \cite{WEI99}; bias dependence of spin-dependent
coherent tunneling \cite{SOU02}; shot noise in the co-tunneling regime \cite{SUK01}; in the Kondo-regime
\cite{MEI02c}. In Ref.~\cite{WAN98a} a numerical investigation
of a spatially extended QD by means of a coherent technique was 
presented. In the current plateau regime the authors find 
a suppressed Fano factor which is  
smaller than one half for symmetric barriers in contradiction to the result of \cite{CHE91}. 
In contrast, an enhanced Fano factor at the current steps was found. 
All of the above references mainly consider the noise due to negative correlations 
(sub-Poissonian noise). 
In the case of negative differential conductance in the current-voltage characteristic,
e.g. in resonant tunneling diodes, positive correlations lead to super-Poissonian noise \cite{IAN98}. 
Furthermore, in capacitively coupled metallic QDs super-Poissonian noise can occur \cite{GAT02}.

Recently, the measurement of the low-frequency shot noise of tunneling through an ensemble of self-organized QDs was 
presented in Ref.~\cite{NAU02} which primarily motivates this work. The current-voltage characteristic for low bias 
is dominated by steps which are presumably due to
tunneling through few QD ground states which are well separated in energy (see also \cite{ITS96,NAR97,HAP99}). 
The corresponding Fano factor shows an average
noise suppression on the current plateau which enables the determination of the effective collector barrier thickness, given
the known
thickness of the emitter barrier \cite{NAU02}. 
At the current steps, Fano factor peaks appear which we have considered theoretically by a master equation approach \cite{KIE03a}.
It was shown that for tunneling through QD states which are not subject to Coulomb interaction, 
Fano factor peaks at the bias position of 
current steps occur, caused by Pauli's exclusion principle.
Good qualitative agreement with experiment was found.

The goal of this paper is the exemplary demonstration of the influence of the Coulomb 
interaction upon the Fano factor 
in a system of two QD states, its interplay with Pauli's exclusion principle, and the 
consequences for future noise 
experiments where Coulomb interaction is present. The outline of the paper is as follows: in Sec.~\ref{sec:model} a brief description 
of the 
master equation formalism \cite{BEE91a} and the calculation of spectral power density 
(where we mainly follow the lines of Ref.~\cite{HER93}) is given. Sec.~\ref{sec:coulomb} contains 
the results of the bias dependent
Fano factor for varying Coulomb interaction energy and in Sec.~\ref{sec:super} the super-Poissonian 
shot noise related
to negative differential conductance will be discussed. All results will be summarized in Sec.~\ref{sec:concl}.


\section{Model}
\label{sec:model}

\begin{figure}[htb]
  \begin{center}
    \includegraphics[width=.3\textwidth]{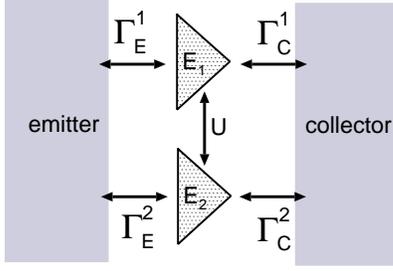}
    \caption{Scheme of the tunneling structure with rates $\Gamma_{E/C}^i$ ($i=1,2$) between two QDs and the 
      emitter/collector, respectively; single-particle
      levels $E_{1/2}$, and Coulomb interaction energy $U$}	
    \label{fig1}
  \end{center}
\end{figure}

The spectral power density of the current fluctuations is related to the autocorrelation function of the current by the 
Wiener-Khintchine theorem: 

\begin{eqnarray}
S_{ab}(\omega )=2\int_{-\infty}^{\infty}dt e^{i\omega t}\left (\langle I_a(t)I_b(0)\rangle - \langle I\rangle ^2\right )
\label{eq:wiener}
\end{eqnarray}

where $a,b=E,C$ (emitter/collector) and $\langle\cdot\rangle$ denotes the ensemble average.
A certain state $\nu =(n_1,\dots ,n_N)$ (occupation numbers $n_i\in\{0,1\}$) at a time $t$ of the considered QD system 
with $N$ single-particle states 
is described by the respective occupation probability $P_\nu(t)$.
The single-particle states may correspond to different energy levels in the same QD, or in different QDs.
The time evolution of these occupation probabilities is determined by sequential tunneling of an electron into or out of the 
emitter/collector contact with the tunneling rates $\Gamma_{E/C}^i$ ($i$ labels the single-particle QD state) and can be 
written 
as a master equation in the general form

\begin{eqnarray}
\dot{\mathbf{\underline P}} = \mathbf{\underline{\underline M}}\,\mathbf{\underline P}
\label{eq:master}
\end{eqnarray}

Note that this approach holds only for weakly coupled QDs, $\Gamma_{E/C}^i\ll k_BT$, and cannot account for 
co-tunneling processes, i.e. coherent tunneling of two electrons simultanously \cite{SUK01,GRA92a}. 
In this paper we consider two QD states $E_{1/2}$ in two different QDs which are connected to the emitter/collector 
contact and are coupled electrostatically by Coulomb interaction of strength $U$ (see Fig.~\ref{fig1}). Then the vector 
of the occupation probabilities is $\mathbf{\underline P}=(P_{(0,0)},P_{(1,0)},P_{(0,1)},P_{(1,1)})^T$ and the transition 
matrix in 
(\ref{eq:master}) reads \cite{BEE91a} 

\begin{widetext}
\begin{eqnarray}
\mathbf{\underline{\underline{M}}}=
\left(\begin{array}{cccc}
-\Gamma_E^1f_E^1-\Gamma_E^2f_E^2  & \Gamma_E^1(1-f_E^1)+\Gamma_C^1 & 
\Gamma_E^2(1-f_E^2)+\Gamma_C^2 & 0 \\
\Gamma_E^1f_E^1 & -\Gamma_E^1(1-f_E^1)-\Gamma_C^1-\Gamma_E^2f_E^{2,U} & 0 & 
\Gamma_E^{2}(1-f_E^{2,U})+\Gamma_C^2 \\
\Gamma_E^2f_E^2 & 0 & -\Gamma_E^1f_E^{1,U}-\Gamma_E^2(1-f_E^2)-\Gamma_C^2 &
\Gamma_E^{1}(1-f_E^{1,U})+\Gamma_C^1 \\
0 & \Gamma_E^2f_E^{2,U} & \Gamma_E^1f_E^{1,U} & \Gamma_E^1f_E^{1,U}+\Gamma_E^2f_E^{2,U}-\Gamma 
\end{array}\right )
\label{eq:matrix}
\end{eqnarray}
\end{widetext}

with $\Gamma:=\Gamma_C^1+\Gamma_C^2+\Gamma_E^1+\Gamma_E^2$ and the Fermi functions in the emitter
$f_E^i=(1+\exp{((E_i-e\eta V)/(k_BT))})^{-1}$ and 
$f_E^{i,U}=(1+\exp{((E_i+U-e\eta V)/(k_BT))})^{-1}$.
$V$ is the bias voltage, $\eta V$ is the voltage drop across 
the emitter barrier, and $f_E^{i,U}$ includes the Coulomb interaction energy $U$ of the occupied QDs. 
For $eV\gg k_BT$ we neglect tunneling from the collector into the QDs,
setting the collector occupation probability $f_C^i=f_C^{i,U}=0$. 

The steady state solution of (\ref{eq:master}) can be obtained by

\begin{eqnarray}
\mathbf{\underline{\underline M}}\,\mathbf{\underline P}^0=0
\label{eq:stat_master}
\end{eqnarray}

In order to calculate the current flowing through the system current operators are introduced. The current 
operators at the collector barrier with $f_C^i=f_C^{i,U}=0$ and at the emitter barrier, respectively, 
are defined by

 \begin{eqnarray}
\label{eq:current_op}
&& \mathbf{\underline{\underline j}}_{C} =
 e\left(\begin{array}{cccc}
 0 & \Gamma_C^1 & \Gamma_C^2 & 0\\
 0 & 0 & 0 & \Gamma_C^2\\
 0 & 0 & 0 & \Gamma_C^1\\
 0 & 0 & 0 & 0
\end{array}\right )\\
&&\mathbf{\underline{\underline j}}_{E}=e
\left(\begin{array}{cccc}
0 & -\Gamma_E^1(1-f_E^1) & -\Gamma_E^2(1-f_E^2) & 0\\
\Gamma_E^2f_E^2 & 0 & 0 & -\Gamma_E^1(1-f_E^{1,U})\\
\Gamma_E^1f_E^1 & 0 & 0 & -\Gamma_E^2(1-f_E^{2,U})\\
0 &  \Gamma_E^1f_E^{1,U} & \Gamma_E^2f_E^{2,U} & 0
\end{array}\right ) \nonumber
\end{eqnarray}

so that the stationary mean current reads

\begin{eqnarray}
\langle I\rangle = \sum_\nu[\mathbf{\underline{\underline j}}_{C}\mathbf{\underline P}^0]_\nu=
\sum_\nu[\mathbf{\underline{\underline j}}_{E}\mathbf{\underline P}^0]_\nu
\label{eq:stat_current}
\end{eqnarray}

In the stationary limit the current at the collector barrier equals the mean 
current at the emitter barrier. For the calculation of the 
stationary current in (\ref{eq:stat_current}) the current operators (\ref{eq:current_op}) 
could also be defined in diagonal form
without changing the result (\ref{eq:stat_current}). 
But for the determination of the current-current correlator (see below) the definition
(\ref{eq:current_op}) becomes crucial as it projects the occupation
  probability to the state after an electron traversed the barrier.

To define the autocorrelation function of the current the time propagator $\mathbf{\underline{\underline T}}(t)$
is introduced as follows:

\begin{eqnarray}
\mathbf{\underline{\underline T}}(t)\equiv\exp{(\mathbf{\underline{\underline M}}t)}\quad\textrm{with}\quad
\mathbf{\underline P}(t)= \mathbf{\underline{\underline T}}(t)\,\mathbf{\underline P}(0)
\label{eq:time}
\end{eqnarray}

With (\ref{eq:stat_master}),(\ref{eq:current_op}), and (\ref{eq:time}) the current-current correlator in (\ref{eq:wiener})
is \cite{HER93}

\begin{eqnarray}
\langle I_a(t)I_b(0)\rangle &=& \theta (t)\sum_\nu[\mathbf{\underline{\underline j}}_{a}\,
\mathbf{\underline{\underline T}}(t)\,\mathbf{\underline{\underline j}}_{b}\mathbf{\underline P}^0]_\nu+\nonumber\\
&&+\theta (-t)\sum_\nu[\mathbf{\underline{\underline j}}_{b}\,
\mathbf{\underline{\underline T}}(-t)\,\mathbf{\underline{\underline j}}_{a}\mathbf{\underline P}^0]_\nu+\nonumber\\
&&+e\delta_{ab}\delta (t)\sum_\nu\left\vert [\mathbf{\underline{\underline j}}_{a/b}
\mathbf{\underline P}^0 ]_\nu\right\vert
\label{eq:correl}
\end{eqnarray}

($\theta (t)$ is the Heaviside function).
The first two terms of the right-hand side in (\ref{eq:correl}) contain the correlation between tunneling events at different 
times. The last term describes the self-correlation of a tunneling event at the same barrier
(for further discussions of (\ref{eq:correl}) see \cite{DAV92,HER93,KOR94}).

For $\omega\ll \Gamma_{E/C}^i$ ($i=1,2$) which is the regime where experimental data are available 
at the moment (e.g. \cite{NAU02}), 
the spectral power densities become constant and
$S_{EE}(0)=S_{CC}(0)=S_{EC}(0)=S_{CE}(0)\equiv S(0)$ holds.

As a measure of deviation from the uncorrelated Poissonian noise the dimensionless Fano factor $\alpha$ is used 
\cite{BLA00}:  

\begin{eqnarray}
\label{eq:fano}
\alpha (0)&\equiv& \frac{S(0)}{S_P}\\
\textrm{with}\quad S_P&=&2e\langle I\rangle\quad\textrm{(Poissonian noise \cite{SCH18})}\nonumber
\end{eqnarray}


\section{Coulomb interacting quantum dots}
\label{sec:coulomb}

We consider the tunneling through two non-degenerate QD ground states $E_1$ 
and $E_2$ with an energy 
separation $\Delta E=E_2-E_1$ which could correspond to slightly different QD sizes. 
The coupling to the emitter and collector contact
will be assumed to be the same for both states: $\Gamma_E\equiv\Gamma_E^1=\Gamma_E^2$ and 
$\Gamma_C\equiv\Gamma_C^1=\Gamma_C^2$.

In the following we discuss three cases for the Coulomb interaction energy $U$: 
{\bf A.} $U=0$, {\bf B.} $U<\Delta E$, and {\bf C.} $U>\Delta E$.


\subsection{Noninteracting states: $U=0$}

In Figs.~\ref{fig2} and \ref{fig3} the results of a calculation for variation of $U$ in the range of a few $k_BT$ 
are shown (for fixed 
$k_BT=\Delta E$/23 and $\gamma =$5). 
The mean current $\langle I\rangle$ vs. bias voltage $V$ is plotted in Fig.~\ref{fig2}a. For $U=0$ there are two steps due to
tunneling through the respective states. 
The width of the current steps is determined by the Fermi distribution of the emitter
electrons. Note that typical energy scales are as follows: the bias voltage $V$ is of the order of tens 
of mV, 
$\Delta E$ can be of the order of a few meV, $k_BT$ is  of the order of tens of $\mu$eV for temperatures in the range of a few Kelvins.

The respective Fano factor $\alpha$ (\ref{eq:fano}) is shown in Fig.~\ref{fig2}b.
On the first plateau in the current-voltage curve of Fig.~\ref{fig2}a where only tunneling through one single-particle state
occurs the Fano factor becomes

\begin{eqnarray}
\label{eq:fano_single}
\alpha_i\equiv\alpha_i(0)=1-\frac{2}{\gamma_i+2+\frac{1}{\gamma_i}}f_E^i\\
\textrm{with}\quad\gamma_i\equiv\frac{\Gamma_C^i}{\Gamma_E^i}\nonumber
\end{eqnarray}

where $i=1$ denotes the energetically lowest state.
For $f_E^1=1$ eq.~(\ref{eq:fano_single}) is the well-known relation derived by L. Y. Chen {\it et al.} \cite{CHE91}. 
It reflects
the sensitivity of the Fano factor to Pauli's exclusion principle. This is shown by the full curve $\alpha(\gamma )$ 
in Fig.~\ref{fig4}:
For symmetric tunneling barriers $\alpha $ is equal to one half and approaches unity for strong asymmetry. 
For bias voltages below the current onset where $f_E^1\approx 0$ the tunneling current becomes uncorrelated so 
that $\alpha_1=1$. 
At the second step where the second single-particle state is filled
the Fano factor (Fig.~\ref{fig2}b) has a peak which is
also an effect of Pauli's exclusion principle and we obtain a simple analytical
expression for an arbitrary number of noninteracting QD states (for a derivation for two states see Appendix A):

\begin{eqnarray}
\label{eq:fano_noninter}
\alpha=\frac{\sum_i\langle I_i\rangle\alpha_i}{\langle I\rangle}
\end{eqnarray}

by (\ref{eq:fano_single}), where the current through state $i$ is: $\langle I_i\rangle =e\frac{\Gamma_C^i}{1+\gamma_i}f_E^i$, 
and the net current is $\langle I\rangle =\sum_i \langle I_i\rangle$.
Eq.~(\ref{eq:fano_noninter}) was applied to the measured Fano factor modulation of tunneling through self-organized QDs 
\cite{NAU02} in a bias regime where only a few QD ground states are active in transport. It can qualitatively reproduce  the
measured Fano factor dependence upon the bias voltage \cite{KIE03a}.


\subsection{$U<\Delta E$}

\begin{figure}[htb]
  \begin{center}
    \includegraphics[width=.45\textwidth]{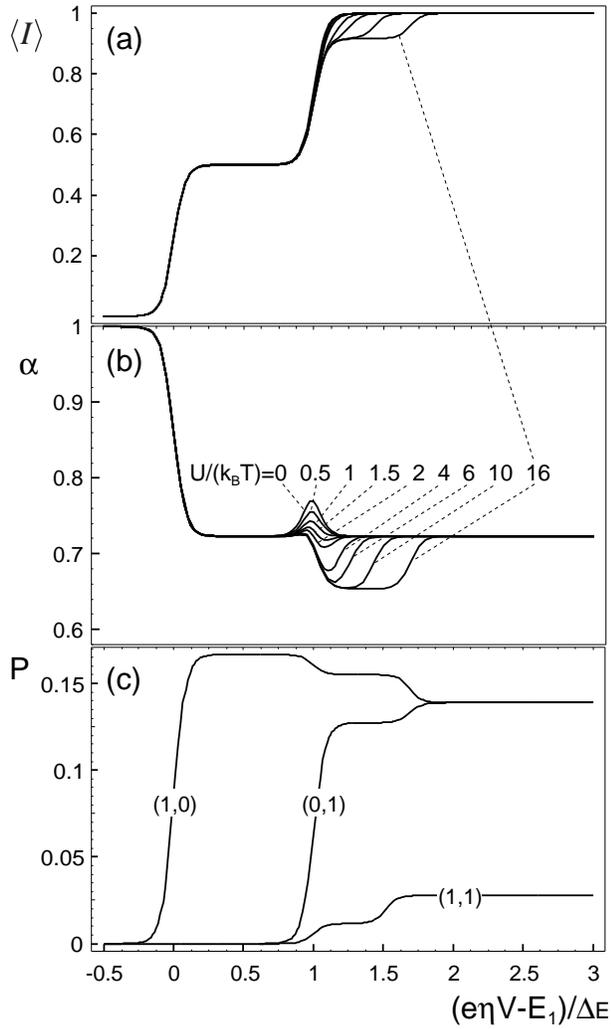}
    \caption{a) Normalized mean current $\langle I\rangle$ vs. bias voltage $V$. b) Fano factor $\alpha$ vs. bias voltage $V$
      for different values of the Coulomb interaction energy $U$. 
      c) Occupation probabilities $P_{(1,0)}$, 
      $P_{(0,1)}$, and $P_{(1,1)}$ vs. bias voltage $V$ for $U/k_BT=$16. Parameters: 
	$k_BT=\Delta E$/23, $\gamma =\Gamma_C/\Gamma_E=$5.}	
    \label{fig2}
  \end{center}
\end{figure}

\begin{figure}[htb]
  \begin{center}
    \includegraphics[width=.45\textwidth]{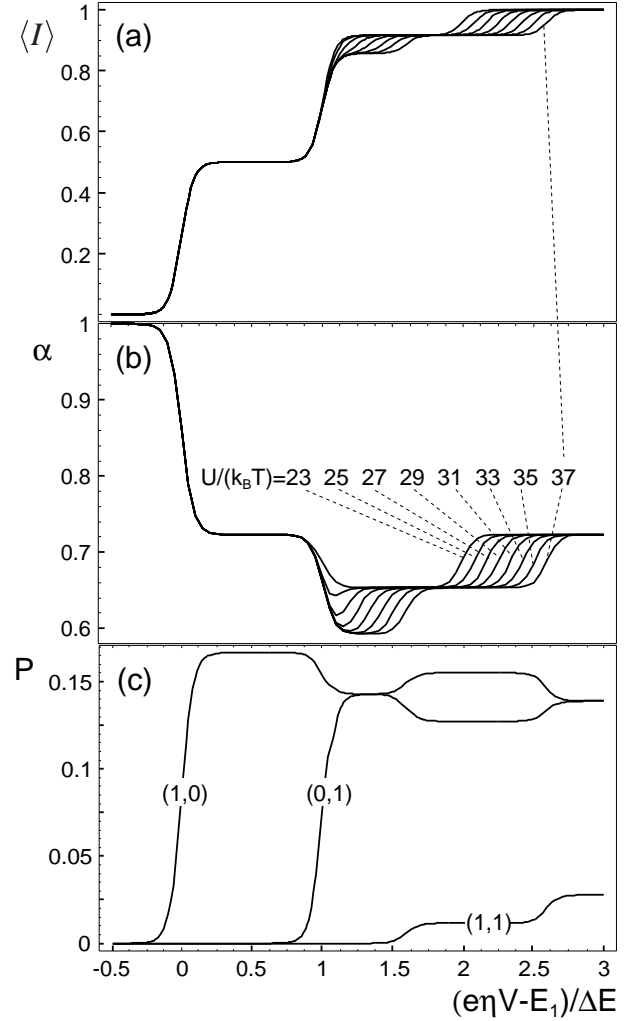}
    \caption{Same as Fig.~\ref{fig2}, but for different values of $U$. 
      c) $U/k_BT=$37.}	
    \label{fig3}
  \end{center}
\end{figure}

With increasing Coulomb interaction $U\neq 0$ the Fano factor peak vanishes (see Fig.~\ref{fig2}b)
while the current changes only slightly. This underlines again the strong sensitivity of shot noise to correlations.

Further increase of $U$ leads to an additional step whose bias voltage is proportional to $U$. 
The respective
occupation probabilities $P_{(1,0)}$, $P_{(0,1)}$, and $P_{(1,1)}$ for $U=16k_BT$ 
are shown in Fig.~\ref{fig2}c:
at the first plateau the electrons tunnel through the energetically lowest state $(1,0)$; the second plateau is generated by 
tunneling through both single-particle states with different probabilities and with lower probability through the 
two-particle state which is determined by the coupling to the collector. 
This correlated state originates from aligning the emitter Fermi energy
with the energy $E_1+\Delta E$ of the second single-particle state which can be filled then. 
If the system is in the state $(0,1)$, a second electron may enter the $i=1$ level,
as $U<\Delta E$. In contrast, the level $i=2$ is not accessible from the state
$(1,0)$ as long as $\eta V<E_1+\Delta E+U$. This explains the asymmetry
between the occupation probabilities $P_{(1,0)}$ and $P_{(0,1)}$.
The height of this second plateau depends on the ratio of the tunneling rates: 
$\propto (1+\gamma)^{-1}$ with $\gamma :=\Gamma_C/\Gamma_E$.

\begin{figure}[htb]
  \begin{center}
    \includegraphics[width=.45\textwidth]{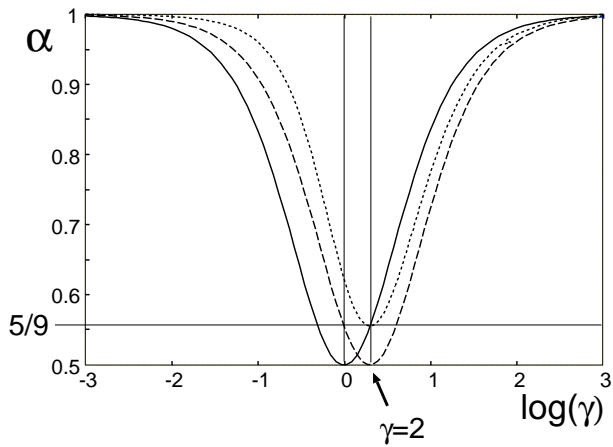}
    \caption{Fano factor $\alpha$ vs. ratio of tunneling rates $\gamma =\Gamma_C/\Gamma_E$. Full curve: first 
      current plateau in Fig.~\ref{fig3}; Dashed curve: second current plateau; 
      Dotted curve: third current plateau. $U=37k_BT$.}	
  \label{fig4}
  \end{center}
\end{figure}

At the second plateau in the current-voltage characteristic the Fano factor differs from the case of $U=0$, 
where only Pauli's exclusion principle plays a role.
The dependence on the ratio of tunneling rates $\gamma$ is shown by the
dotted curve in Fig.~\ref{fig4}. In contrast to the noninteracting regime the minimum is now at $\gamma =2$ with 
$\alpha =5/9$. These are numerical values, as we did not obtain an analytical expression
for this Coulomb correlated state.

\begin{figure}[htb]
  \begin{center}
    \includegraphics[width=.45\textwidth]{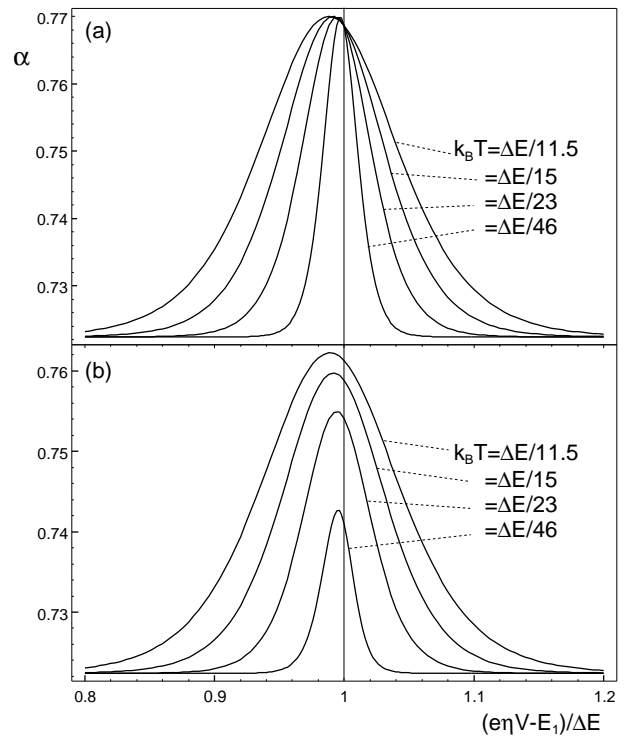}
    \caption{Temperature dependence of the Fano factor vs. bias voltage $V$ for $\gamma =$5. 
	a) $U=0$. b) $U=\Delta E$/46}	
    \label{fig5}
  \end{center}
\end{figure}

In the experiment of Ref.~\cite{NAU02} the question arises whether the QD states which are contributing to
transport are Coulomb interacting. One way of determining this question would be the analysis of the Fano factor 
dependence upon the tunneling rate ratio
$\gamma$ as shown in Fig.~\ref{fig4}. However, in the experimental setup of  Ref.~\cite{NAU02} these rates are determined
by the growth procedure. Therefore, they cannot be varied in the same sample. 
Here, we propose how to obtain the information about Coulomb correlations via the temperature dependence of the
Fano factor. In Fig.~\ref{fig5} the Fano factor $\alpha$ vs. bias voltage $V$ for different temperatures $T$ in the bias 
range of the second current step of Fig.~\ref{fig2} is plotted. For noninteracting QD states (Fig.~\ref{fig5}a) 
the Fano factor peak gets broader and experiences a slight shift to lower bias voltages for increasing temperatures.
A qualitatively different picture results for interacting QD states in Fig.~\ref{fig5}b ($U=\Delta E/46$): with 
increasing temperature
the peak increases and also shifts to lower voltages. Hence, a unique fingerprint of Coulomb interaction of QD states 
even for very small $U$ shows up in the temperature dependence of the Fano factor peaks.
Experimental investigations of this are in progress \cite{NAU02a}.
 
\subsection{$U>\Delta E$}

For $U>\Delta E$ a fourth step arises in the current vs. bias voltage characteristic in Fig.~\ref{fig3}a). Now, the
second current plateau corresponds to a different state as in section B. Due to $U>\Delta E$ only the two single-particle
states can be filled with the same probability (compare Fig.~\ref{fig3}c). The respective Fano factor dependence
upon $\gamma$ is (as discussed in Ref.~\cite{NAZ96})   

\begin{eqnarray}
\alpha=1-\frac{2\tilde{\Gamma}_E\Gamma_C}{(\tilde{\Gamma}_E+\Gamma_C)^2}
\quad\textrm{with}\quad \tilde{\Gamma}_E = 2 \Gamma_E
\label{eq:fano_correl}
\end{eqnarray}

and is shown by the dashed curve in Fig.~\ref{fig4}. The effect of Coulomb correlation upon the Fano factor consists 
substituting
$\Gamma_E$ by $2\Gamma_E$ in eq.~(\ref{eq:fano_single}) and leads to a shift of the minimum of the full curve 
in Fig.~\ref{fig4} by $\gamma =2$. For $\gamma =$1 the Fano factor is $\alpha =$5/9.

   
\section{Super-Poissonian noise}
\label{sec:super}

\begin{figure}[t]
  \begin{center}
    \includegraphics[width=.45\textwidth]{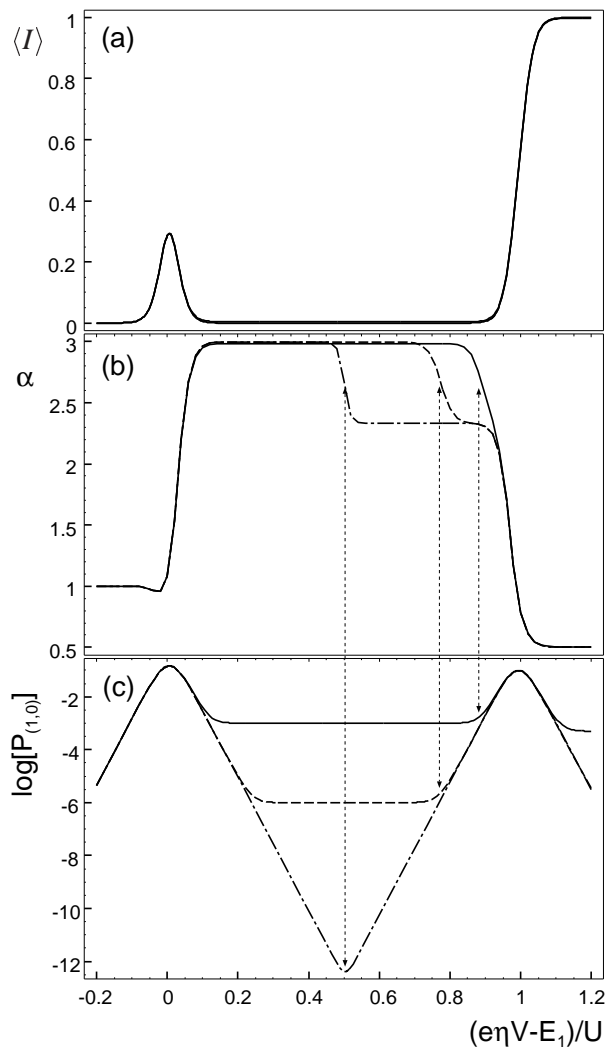}
    \caption{a) Normalized mean current $\langle I\rangle$ vs. bias voltage $V$. b) Fano factor $\alpha$ 
      vs. bias voltage $V$. c) Occupation probability $P_{(1,0)}$ 
      vs. bias voltage (logarithmic scale) 
      for $\gamma_2=0$ (dash-dotted curve), $\gamma_2=10^{-6}$ (dashed curve), $\gamma_2=0.001$ (full curve).
      Parameters: $\gamma_1=1$, $k_BT=U$/23, $\Delta E=$0. Note that all three curves coincide in (a).}	
    \label{fig6}
  \end{center}
\end{figure}

\begin{figure}[t]
  \begin{center}
    \includegraphics[width=.45\textwidth]{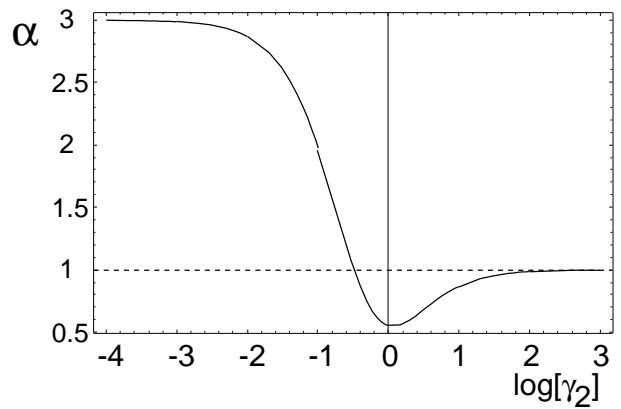}
    \caption{Fano factor $\alpha$ vs. $\gamma_2$ at $(e\eta V-E_1)/U=$0.2 in Fig.~\ref{fig6}b.}	
    \label{fig7}
  \end{center}
\end{figure}

In the previous section we used identical rates $\Gamma_{C/E}$ for both states. Now, we allow their couplings to the collector
contact to be different: $\Gamma_C^1\neq\Gamma_C^2$. For $\gamma_2\leq 0.45\gamma_1$ 
this leads to negative differential 
conductance (NDC) in the current-voltage characteristic as shown in Fig.~\ref{fig6}a ($\Delta E=0$) \cite{KIE02}. 
If the tunneling rate
to the collector of state $i=2$ is much smaller than the other one the occupation probability $P_{(0,1)}$ becomes close to unity. 
The 
current carried by this state is proportional to $\Gamma_C^2$ and therefore low. Consequently,
the occupation probability $P_{(1,0)}$ is low except close to the current onset where the respective current peak arises 
and around the bias voltage where the 
two-particle state $(1,1)$ becomes occupied.
(compare Fig.~\ref{fig6}c). Let us consider the dependence of the Fano factor on the bias voltage in Fig.~\ref{fig6}b: in
the bias voltage range where the current is suppressed the Fano factor is larger than unity
(super-Poissonian shot noise caused by positive correlations of tunneling events). This is similar to the situation in a 
resonant tunneling diode \cite{IAN98}. 
The Fano factor exhibits two different values separated by a step
in the middle of the NDC-region (dashed curve in Figs.~\ref{fig6}b) and c)) marked by arrows. 
The Fano factor dependence on $\gamma_2$ in the regime where $P_{(1,0)}$ decreases or keeps constant is shown in 
Fig.~\ref{fig7}. This is the actual NDC regime: for 
$\gamma_2\rightarrow 0$ $\alpha$ approaches 3. In the bias regime where $P_{(1,0)}$ starts to increase due to thermally 
activated
electrons (arrows in Fig.~\ref{fig6}c) the Fano factor is lowered
due to the effect of Pauli's exclusion principle. Its value depends on the coupling ratio $\gamma_1$.


\section{Conclusions}
\label{sec:concl}

We have investigated the low-frequency shot noise behavior in tunneling through two non-degenerate QD states which
interact electrostatically. For noninteracting states the respective non-equilibrium current-voltage characteristic 
shows steps due to resonant tunneling through the single-particle states. In this case we have derived an explicit 
analytical expression
for the bias dependence of the Fano factor. For interacting states additional current steps occur 
which are associated with Coulomb correlated states.
The influence of the couplings 
to the collector upon the Fano factor has been clarified. We have developed sensitive tools to determine
whether states 
are Coulomb correlated in tunneling
experiments through self-organized QDs. This can be done by investigating the temperature dependence of the Fano factor vs.
bias voltage at steps in the current-voltage characteristic.

Furthermore, we have examined an NDC mechanism in a system where the two states are coupled to the collector with different
tunneling rates.
Then the weakly coupled state blocks the other state by the Coulomb interaction. For degenerate states a current peak
occurs and we have analyzed the Fano factor which becomes larger than unity 
(super-Poissonian noise due to positive correlations) 
in the bias range where the current is Coulomb blocked. 
In the limit of vanishing coupling of one state to the collector, we find $\alpha =$3.

\begin{acknowledgments}
The authors would like to thank A. Nauen and R. J. Haug for helpful discussions.
This work was supported
by Deutsche Forschungsgemeinschaft in the framework of Sfb 296.
\end{acknowledgments}


\appendix
\section{Analytical evaluation of the Fano factor for two noninteracting states}

The stationary probability that level $i$ is occupied is $p_i\equiv\frac{\Gamma_E^if_E^i}{\Gamma^i}$ or unoccupied $1-p_i$
($\Gamma^i:=\Gamma_E^i+\Gamma_C^i$).
Then the stationary occupation probability of the noninteracting two-level system given by (\ref{eq:stat_master}) reads 

\begin{eqnarray}
\mathbf{\underline{P}^0}=
\left(\begin{array}{c}
(1-p_1)(1-p_2)\\
p_1(1-p_2)\\
p_2(1-p_1)\\
p_1p_2
\end{array}\right )
\end{eqnarray}

since in the uncorrelated case the occupation probability for each state factorizes into the occupation probabilities of
the single levels. 
By inserting this vector into (\ref{eq:stat_current}) one immediately sees that terms with $p_ip_j$ cancel
and the current is the sum of the currents through each level $i$: $\langle I_i\rangle =e\Gamma_C^ip_i$. 
This also holds for an arbitrary number of levels: $\langle I\rangle = \sum_i\langle I_i\rangle$.

Now, let us consider the time propagator (\ref{eq:time}): its matrix elements $T_{\nu\mu}(t)$ describe the conditional
probability to have state $\nu$ at time $t$ under the condition of state $\mu$ at $t=0$.
The matrix element $T_{\nu\mu}(t)\equiv T_{\mu\rightarrow\nu}(t)$ can be factorized for each level $i$ 
with the following conditional
probabilities:

\begin{eqnarray} 
 n_i&=&0\rightarrow 1:\quad p_i(1-e^{-\Gamma^it})\nonumber\\
 n_i&=&1\rightarrow 0:\quad (1-p_i)(1-e^{-\Gamma^it})\nonumber\\
 n_i&=&0\rightarrow 0:\quad 1-p_i(1-e^{-\Gamma^it})\nonumber\\
 n_i&=&1\rightarrow 1:\quad p_i+e^{-\Gamma^it}(1-p_i)
\label{eq:cond_prob}
\end{eqnarray}

Due to the form of the current operator at the collector barrier in (\ref{eq:current_op}) the first row and last column 
of the matrix 
$\mathbf{\underline{\underline T}}(t)$ does not enter in the calculation of the current-current correlator 
(\ref{eq:correl}). Carrying out the sum in (\ref{eq:correl}) for two levels leads to

\begin{widetext}
\begin{align}
\langle I_C(t)I_C(0)\rangle = &\ 2e\Gamma_C^1\left\{ [\langle I_1\rangle p_1(1-p_2)+\langle I_2\rangle p_2(1-p_1)]
[T_{(0,0)\rightarrow (1,0)}+T_{(0,0)\rightarrow (1,1)}]\right.\nonumber\\
&\left.+\langle I_2\rangle p_1[T_{(1,0)\rightarrow (1,0)}
+T_{(1,0)\rightarrow (1,1)}]+\langle I_1\rangle p_2[T_{(0,1)\rightarrow (1,0)}+T_{(0,1)\rightarrow (1,1)}]\right\}\nonumber\\
&+2e\Gamma_C^2\left\{[\langle I_1\rangle p_1(1-p_2)+\langle I_2\rangle p_2(1-p_1)]
[T_{(0,0)\rightarrow (0,1)}+T_{(0,0)\rightarrow (1,1)}]\right.\nonumber\\
&\left.+\langle I_2\rangle p_1[T_{(1,0)\rightarrow (0,1)}+T_{(1,0)\rightarrow (1,1)}]
+\langle I_1\rangle p_2[T_{(0,1)\rightarrow (0,1)}+T_{(0,1)\rightarrow (1,1)}]\right\}\nonumber\\
&+e\langle I\rangle \delta (t)
\label{eq:correl_expl1}
\end{align}
\end{widetext}

Replacing the $T_{\mu\rightarrow\nu}$ in (\ref{eq:correl_expl1}) by using the rules (\ref{eq:cond_prob})  the correlator
becomes

\begin{align}
\langle I_C(t)I_C(0)\rangle =&\ -2\langle I_1\rangle^2e^{-\Gamma^1t}-2\langle I_2\rangle^2e^{-\Gamma^2t}+\langle I\rangle^2\nonumber\\
&+e\langle I\rangle \delta (t)
\label{eq:correl_expl2}
\end{align}

which can be generalized for an arbitrary number of levels

\begin{align}
\langle I_C(t)I_C(0)\rangle =&\ -2\sum_i \langle I_i\rangle^2e^{-\Gamma^it} + \langle I\rangle^2\nonumber\\
&+e\langle I\rangle \delta (t)
\label{eq:correl_expl3}
\end{align}

The time-independent term in (\ref{eq:correl_expl2}) and (\ref{eq:correl_expl3}) cancels out in the calculation of the
spectral power density (\ref{eq:wiener}) and we obtain

\begin{eqnarray}
S(0) = 2e\langle I\rangle - 4 \sum_i\frac{\langle I_i\rangle^2}{\Gamma^i}
\label{eq:power}
\end{eqnarray}

Dividing Eq.~(\ref{eq:power}) by $2e\langle I\rangle$ and using the Fano factor for tunneling through a single level $i$, i.e.
$\alpha_i=1-\frac{2\langle I_i\rangle}{e\Gamma^i}$ (\ref{eq:fano_single}), 
the Fano factor for an arbitrary number of noninteracting
levels eq.~(\ref{eq:fano_noninter}) is derived.


\end{document}